\documentclass{nime-alternate}

\begin{document}
%
\conferenceinfo{NIME'15,}{May 31-June 3, 2015, Louisiana State Univ., Baton Rouge, LA.}

\title{Enacting Musical Worlds: Common Approaches to
  using NIMEs within Performance and Person-Centred Arts Practices}

%
%
%
%
%

\numberofauthors{1} 
%
\author{
%
%
\alignauthor
{Lauren Hayes}\\
       \affaddr{Arts, Media + Engineering}\\
       \affaddr{Arizona State University}\\
       \affaddr{Tempe, AZ 85287}\\
       \email{laurensarahhayes@gmail.com}
}

\maketitle
\begin{abstract}
Live music making can be understood as an enactive process, whereby
musical experiences are created through human action. This suggests
that musical worlds coevolve with their agents through repeated
sensorimotor interactions with the environment (where the music is
being created), and at the same time cannot be separated from their
sociocultural contexts. This paper investigates this claim by
exploring ways in which technology, physiology, and context are bound
up within two different musical scenarios: live electronic
musical performance; and person-centred arts applications of NIMEs.

In this paper I outline an ethnographic and phenomenological enquiry into
my experiences as both a performer of live electronic and
electro-instrumental music, as well as my extensive background in
working with new technologies in various therapeutic and
person-centred artistic situations. This
is in order to explore the sociocultural and technological contexts in
which these activities take place. I propose that by understanding
creative musical participation as a highly contextualised practice,
we may discover that the greatest impact of rapidly developing
technological resources is their ability to afford richly diverse,
personalised, and embodied forms of music making. I argue that this
is applicable over a wide range of musical communities.
\end{abstract}


\acmclassification1998{

H.5.5 [Information Interfaces and Presentation] Sound and Music
Computing---Methodologies and techniques, 
J.5 [Arts and Humanities] Music}

\section{Introduction}
Christopher Small's concept of musicking firmly places participation
at the centre of what it is \emph{to music}. To take part in a musical
activity---which includes sweeping the stage before a concert, selling
tickets, in addition to accepted musical practices such as composing
or performing---entails the forging of various relationships. Small
argues that it is through these relationships, which may exist between
people, sounds, and spaces, that meaning is constructed. 

In what follows I provide two contrasting accounts of NIME
development. Through these, I explore how the relationships that Small
outlines are forged over time through the lens of practice-led and
ethnographic research. I offer examples which lie within two often
unrelated areas: the Western experimental and electronic music
communities; and the world of person-centred arts practices for people with
complex disabilities. This is in order to
illustrate how such an understanding of NIME development within one
context may inform work in another. This may also illuminate parallels
between what could be perceived as unrelated musical practices.

By viewing musical engagement as an evolving and embodied process,
which supports Small's definition of music as human action, it can be
demonstrated that the relevance of technological developments in the
field of live electronic and digital musical practice lies not
necessarily within the material aspects per se. But rather, an
important consequence is the potential for individualised
practices to emerge, where each performer enacts a unique musical
environment in coordination with their physiological, cultural,
social, and musical histories. I will suggest that by viewing
NIME-related practices in this way, we are afforded the opportunity to
view musical activity \textit{in general} as a ``medium of social
relation'' \cite{DeNora:2000zr} in various contexts.

\subsection{Perceptually Guided Action}
The idea of performance as perceptually guided action
\cite{Hayes:2013fk} suggests the importance of a multimodal approach
to developing digital musical instruments (DMIs) or NIMEs. Moreover,
this can inform our understanding of musical participation in
general. Not only as listeners, but also as performers we are
continuously making use of multiple streams of sensory feedback as we
make our way through a performance: auditory, haptic, kinaesthetic,
and visual.

This draws on Francisco Varela, Evan Thompson and Eleanor Rosch's
theory of enaction \cite{Varela:1991fk} as a way of understanding the
importance of the role of the body within---specifically---live
electronic musical performance. An enactive understanding focuses on
the idea of structural coupling between agent and environment through repeated
sensorimotor interactions \cite{Varela:1991fk}. Both the perceptive capacities of various
organisms as well as the environment itself emerge through reciprocal
coordination and coevolution. In biological terms, for example, this
phenomenon is ``responsible for both the ultraviolet vision of bees
and the ultraviolet reflectance patterns of flowers''
\cite{Varela:1991fk}. Through phenomenological enquiry we start to see
how musical worlds may evolve in a similar manner.

The concept of enaction extends Maurice Merleau-Ponty's work on
phenomenology, which posits the body as both the perceiving object,
and at the same time, the subject of perception
\cite{Merleau-Ponty:1962uq}. Merleau-Ponty illustrates the body's
capacity for this duplicity of sensation through an example of the
hands, which oscillate between touching and being touched.

The enactive approach emphasises the mutuality between agent and
environment. Similarly, musical works can emerge out of the relationships
that develop over time between a specific combination of people,
instruments, and space. This does not apply only to the immediacy of musical
performance. This framework can also be used to understand the
durational development of NIMEs, where an instrument may be iterated
through a series of incremental adjustments informed by the experience
of their use within different scenarios. 

Small suggests that when we rehearse and
perform, we are exploring not only the sonic relationships that
articulate how we ideally believe sounds should be organised, but also
the relationships between sound and instruments; the relationship between the
performers and the audience; the relationship between those taking
part and the physical setting; and so on [1].

\subsection{Ethnography and Creative Practice} 
There has been a growing number of calls from the NIME community to
acknowledge the importance of both ethnographic (see \cite{Booth2012})
and practice-led (see \cite{green2014nk}) research. These
methodologies allow for a discussion of the complex relationships
between the sociocultural contexts in which technical developments in
various NIME-related fields are being made. Both creative practice and
ethnographic approaches provide space for exploring how NIME-related
research unfolds over time in the real world. The two case studies
that I describe each offer accounts of highly personalised NIME
development. In exploring these situations, it will become apparent
that any attempt to optimise the instruments discussed for the wider
community would be largely redundant as they are evolved through the
physiologies and aesthetic choices of the specific musicians
involved. Evaluation of these practices through objective testing
would be fruitless. The methodologies employed allow the experiences
of the users to be shared through reflective observation and
discussion. Nevertheless, by offering this insight, we can start to
find implications for engagement with DMIs in general.

\section{Two Case Studies}
In this section I discuss two case studies where I have developed
NIMEs in what initially appears to be unrelated contexts: my own live
electronic performance practice; and person-centred collaborative arts
practice. In
each case I examine the role of the physiology of the musician, their
musical aesthetics and histories, the contexts in which the musical
engagement takes place, and how this is bound up with the various
technologies employed.

\subsection{Personal Performance Practice}
\subsubsection{Background}
Over the last eight years, I have explored an approach to personal DMI
design that focuses around the relationships between sound and
touch. This explores the double aspects of Merleau-Ponty's notion of
embodiment through, on one hand, themes of resistance and haptic
technology \cite{Hayes2012kl}, but also the perception of sound as
vibration, through vibrotactile technology \cite{Hayes:2011fk}. While
the benefits of using haptic technology for improving certain aspects
of instrumental skill acquisition are well documented \cite{silephd},
research in this area tends to be focused around technical
development. My own research has attempted to provide an in-depth,
practice-based perspective in this field.

\subsubsection{The Physiology of the Performer}
It is perhaps not surprising that my training as a classical pianist,
which began formally at the age of four, has led to an exploration of
musical HCI that is largely focused around the expressive capacities
of the fingers. While I may have been drawn to the piano simply due to
its ubiquity as a traditional Western instrument, through repeated
engagement with the instrument from this young age, by way of lessons,
exercises, and the sort of experimentation that I much later learned
was called improvisation, I enacted my musical environment based
around a very specific type of tactility.

I learned to make use of both the vibrotactile feedback of the
resonating body of the piano, as well as the particular resistances that
it offered me as a physical instrument. When much later I started to
perform with computers, the disconnect between sound and touch left me
unfulfilled as a musician. Performance gestures contained none of the
effort, struggle, or physicality that I was accustomed to making use
of. 

This led me to question what it was that I was missing in my
experience as a performer now engaging with digital technology, in
order to adequately communicate a musical idea. How could I translate
an intention into an expressive and articulated sonic result? It has
been through my own personal history of musical performance that I
have been prompted to examine the relationship between sound and
touch. Experiencing what Simon Emmerson describes as ``increasingly
alienated from purely physical sound production'' \cite{emmersontouch},
urged me to explore more deeply the links between action and
perception, specifically for the performer.

This research has been extensively documented elsewhere, and has
included incorporating haptic technologies \cite{Hayes2012kl,
Hayes:2013fk}, and vibrotactile feedback \cite{Hayes:2011fk} into my
instrument design. Rather than reiterate the technicalities of this work,
it is important to note that my evolved means of musical
expression has been closely tied to my physiology over a long period
of time. As I approached NIME development, this relationship between
body and instrument was key to informing my design choices in that
they were deeply rooted within an exploration of physicality and touch.

\subsubsection{Sociocultural Context}
My performance practice using NIMEs has been largely situated under
the umbrella of Western art music. Being based within universities has
provided me with access to expertise in both software and hardware
development, situated me within a community of potential
mentors and collaborators, and offered me many opportunities to
present and discuss my work.

Working in and between genres such as contemporary classical, free
improvisation, experimental beat-based music, and noise, has allowed
my various performance environments to be explored within a broad
range of scenarios. For example, at an academic conference, NIMEs used
for performance must be extremely reliable and stable, with the
ability to be set up quickly as there is often little time for this
between performances. Within improvisation scenarios, the NIME must be
flexible and adaptable. It must be able to give space to
co-performers, yet possess a voice of its own. At a noise gig, my
digital/laptop-based instruments must be able to hold their own in
terms of sonic depth against analogue counterparts from myself or
collaborators.


\subsubsection{Objectives for NIME Development}
While a large part of my practice has been based around the hybrid
piano, formed around haptically and digitally augmented acoustic
pianos \cite{Hayes:2013fk}, I have also performed extensively using
a variety of hybrid (analogue and digital) electronic systems. These
are assemblages of various components, including analogue
synthesisers, hardware drum machines, various MIDI and game
controllers, foot pedals, and bespoke software built using
Max/MSP. 

\begin{figure}[htbp]
	\centering
		\includegraphics[width=1\columnwidth]{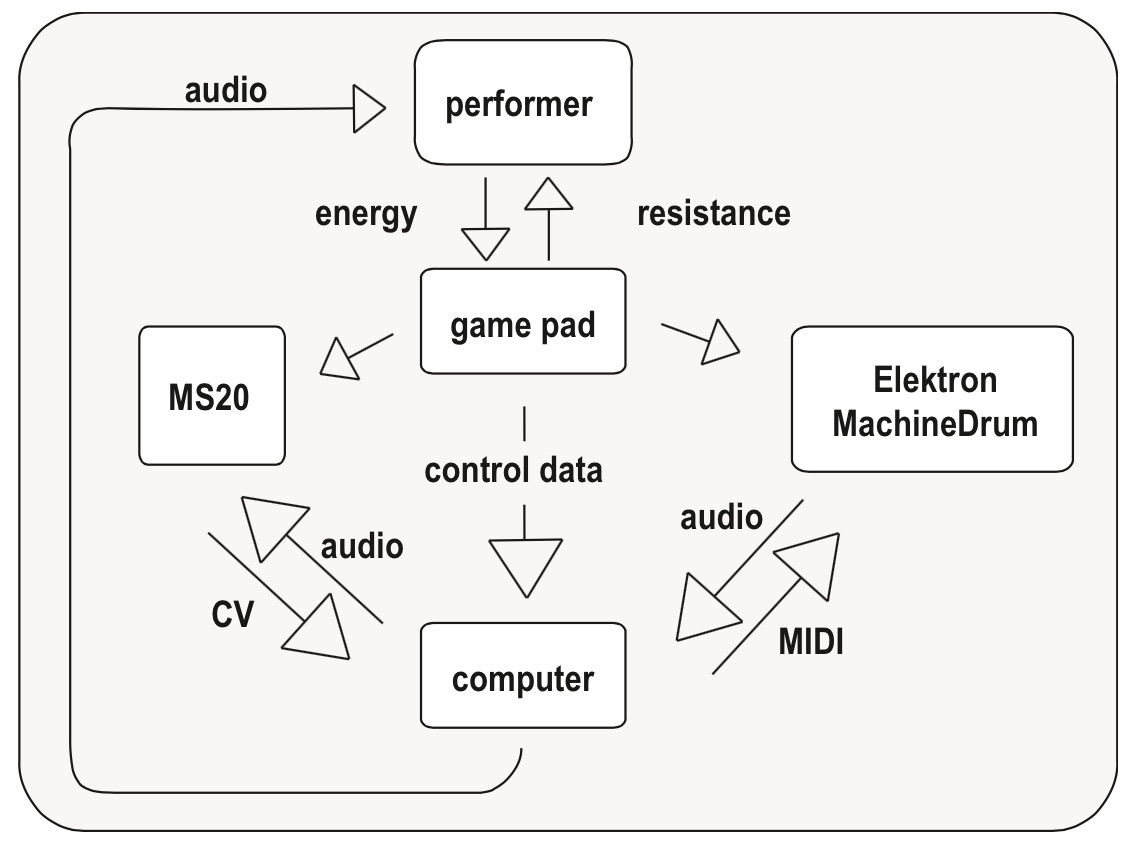}
	\caption{Networked interactions.}
	\label{fig:BlockDiagram1}
\end{figure}

In working with hardware that does not offer the same rich physicality
of the hybrid piano, I had to develop ways of introducing this tactile
engagement to my performance environments. By interacting with all
elements of a particular set up through a single game controller, I
was able to simultaneously touch and engage with different parts of
the instrument, bringing a sense of immediacy into my hands. For
example, in one configuration, the game controller would trigger very
short segments of sounds, which were in turn analysed by the
software. This would send both MIDI information out via the sound card
to trigger drum machine synthesis, as well as sending multiple control
voltages out to an analogue synthesizer. At the same time, the audio
output from these two external devices was sent back into the
laptop. This would then be sampled and processed in Max/MSP, where
several parameters were affected by my own interaction with the game
controller. In this way I was able to access several parts of my
performance system at once, bypassing some of the given---and to me,
undesirable---control interfaces, such as the knobs of my Korg MS20
analogue synthesizer, or the buttons of the Elektron MachineDrum.

While this approach involves initial one-to-many mapping choices,
the overall result is a network of interdependent processes, which
feed into each other. The resistances in my performance environments
often lie within the extreme potential for activity through
interconnections within the audio signal path, which must be
negotiated by the performer, often through holding a static position
for extended periods of time. The game controller, for example, is so
easy to manipulate, that the musicality comes from resisting this by
holding both thumbs fixed on the joysticks, which requires a great
deal of pressure from the hands, and creates a tension in the body: a
movement of even one millimetre can drastically alter the sound.

\subsection{Person-Centered Arts Practice}
\subsubsection{Background}
Since 2006 I have worked in various music therapy-related and
person-centred arts practice roles in the
UK. These have included performing classical piano concerts in day
care centres for adults with learning difficulties, as well as running
several series of workshops for people with complex disabilities. In these
workshops I employ a variety of traditional instruments along with
numerous NIMEs, and I focus on the tangible experience of playing and
perceiving sound.

In 2012 I was asked to be involved with the
Artlink\footnote{\url{http://www.artlinkedinburgh.co.uk/}} Ideas Team,
established by the Edinburgh-based arts charity in 2010. This team,
consisting of various local and
international artists and members of Artlink staff, works with an
individual together with psychologists, care workers and family
members. The goal is to establish new ways of thinking about and
making artworks by exploring creative responses to the daily
experiences of people with profound learning disabilities. I was asked to work alongside artist Steve Hollingsworth to
develop an instrument for a young autistic man (M), who was
intensely drawn to piano
performance\footnote{\url{http://issuu.com/artlinkedinburgh/docs/artlink201112}}.

\subsubsection{The Physiology of the Performer}
``The nature from which man has selected his musical styles is not only
external to him; it includes his own nature---his psychophysical
capacities and the ways in which these have been structured by his
experiences of interaction with people and things, which are part of
the adaptive process of maturation in culture'' \cite{Blacking1973}.

Before any part of the design or build commenced, it was crucial to
spend time with M at the piano to observe his engagement with the
instrument. I visited him weekly to hear him play acoustic
pianos as well as my digitally-augmented instrument, the hybrid piano
\cite{Hayes:2013fk}. Several important observations were made over the
weeks that I visited him. Aside from a piano-based instrument being
selected due to M's enthusiasm for it, his often
overpowering strength meant that keyboards or MIDI instruments were
ruled out as they were neither sturdy nor durable enough.

There was a marked difference in M's playing when sitting at the piano
alone, compared to when we played together. When playing alone, M
would hammer the keys for several minutes at a time with much
force. When playing together, he would pause to make eye contact, to
listen, and to respond. He would mimic patterns that my fingers made
on the keys, often loosely, and sometimes with accuracy. This
posed the first question as to how I could achieve this sense of
engagement in M's new instrument without me, or another musician,
being physically present.

Another important factor was the fact that M could get so
enthralled in playing the piano that he would often become
over-excited, and begin to sweat and hyperventilate. The posed a
further problem as to how we could create something that could be
enjoyable, without being over-stimulating.

\subsubsection{Sociocultural Context}
M's fondness for the piano did not stem from a particular interest
in classical or romantic music, or from a childhood that involved
taking piano lessons. The piano for M was a direct means of
expression: tangible and immediate. This unique connection between
player and instrument allowed an approach to developing the NIME that
could view M's aesthetic choices as based around sounds that were
enjoyed by him in his everyday life. Sound libraries were created
which contained samples of his mother's voice, the sound of his dog
barking, as well as car sounds, and sounds from car racing television shows which
he enjoyed. Additionally, I was able to observe during our time
together in workshops which types of sounds from the hybrid piano
seemed to engage M, and which he seemed to dislike. As M had
been improvising with me over several months, I sampled some of my
own piano playing as source material.

\subsubsection{Objectives for NIME Development}
There were several practical design choices that we had to consider
from the outset. As mentioned above, the instrument had to be stable. Furthermore, M was attracted to wires, and would grab
at any that were visible, so everything had to be hidden and
enclosed. A button interface was proposed, where buttons would be
secured onto the front of the piano. These were combined with LED
lights and vibration motors which offered direct sensory feedback to
M to confirm that he had pressed the buttons. The buttons would
change the sample sets, and stop and start the sounds.

\begin{figure}[htbp]
	\centering
		\includegraphics[width=1\columnwidth]{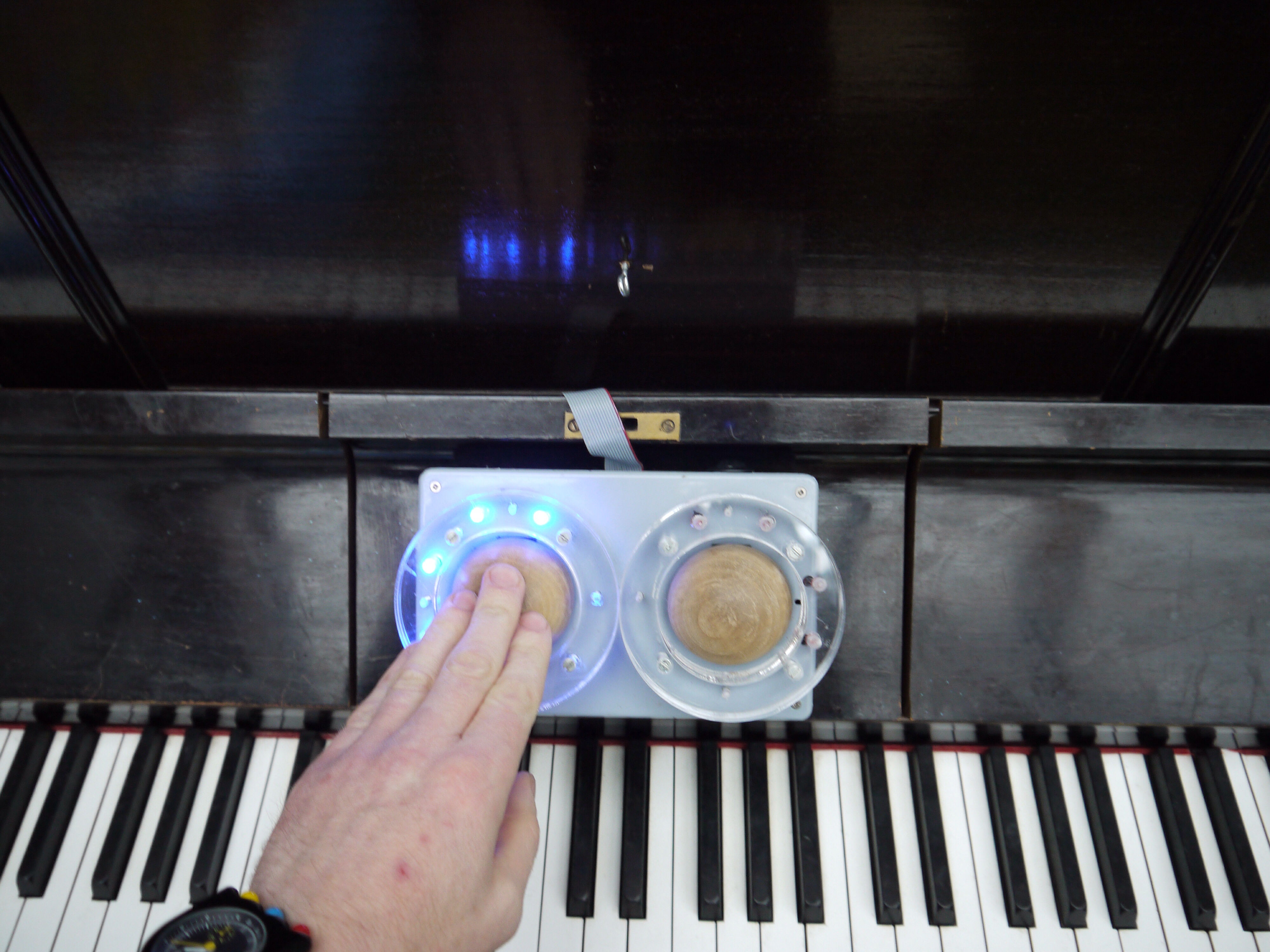}
	\caption{Detail of M's Piano.}
	\label{fig:BlockDiagram2}
\end{figure}

One of the first goals that was discussed in a meeting by M's mother was the idea for M to put a ``square peg in a
square hole''\footnote{Personal email correspondence with Alison
Stirling}. That is to say that the instrument should be able to
demonstrate some degree of agency and intentionality from
M. Furthermore, we envisaged that this would develop over time as
M grew familiar with the instrument. As such, I had to ensure that
the software could be updated when necessary to ensure that it would remain
challenging yet engaging for M. The buttons began as
simple start/stop switches, but could be modified to react differently
if M started to exhibit choice in his actions. The project is ongoing
and the Ideas Team continues to learn from M. 

\section{Conclusion}
I have described two examples of NIME development in differing
contexts. Each situation
uses musical performance as a means of enabling creativity,
expression, communication, and also personal development. The
commonalities between the two situations are clear. In each case, the
development of NIMEs has evolved through close attention to the users'
physiologies and sensorimotor capacities; their musical histories; and
the sociocultural contexts in which the musical engagement takes
place.

It is evident that the type of physical, tactile engagement that I
seek within live electronic musical performance has arisen out of my
perceptive capacities and experience with the touch-based expressivity
of the acoustic piano. However, there is now a generation of musicians
who may not experience this loss of touch since their initial
engagement with music may come through digital devices such as iPads,
rather than acoustic instruments. The model of structural coupling
between musician and instrument that I describe suggests a need for
individualised systems that arise out of the specific interactions of
each person in the world over time. For those with complex
disabilities, an acknowledgement of the uniqueness of their
experiences and responses is required before progress can be made.

If we view music as an enactive process, where new
technologies lead to more engaging, embodied relationships between
people and instruments, both the social as well as the sonic
relationships that we wish to explore, affirm and celebrate [1] can be
realised. The two examples in this paper serve to highlight the need
for more in-depth collaborative research that will combine creative practice
and ethnography with DMI design in order to provide a better
understanding of the experiences and benefits of using new
technologies within musical contexts.

\section{Acknowledgments}
Thanks to Alison Stirling and Kara Christine, along with Steve Hollingsworth, for giving me the opportunity to
work with them on the Artlink Ideas Team.

%
\bibliographystyle{abbrv}
\bibliography{bib}  

\begin{thebibliography}{10}

\bibitem{Blacking1973}
J.~Blacking.
\newblock {\em How Musical Is Man?}
\newblock University of Washington Press, Seattle and London, 1973.

\bibitem{Booth2012}
G.~Booth and M.~Gurevich.
\newblock Collaborative composition and socially constructed instruments:
  Ensemble laptop performance through the lens of ethnography.
\newblock In {\em Proceedings of the 2012 conference on New Interfaces for
  Musical Expression (NIME)}, Michigan, Ann Arbor, 2012.

\bibitem{DeNora:2000zr}
T.~DeNora.
\newblock {\em Music in Everyday Life}.
\newblock Cambridge University Press, Cambridge, 2003.

\bibitem{emmersontouch}
S.~Emmerson.
\newblock `{L}osing {T}ouch?': {T}he {H}uman {P}erformer and {E}lectronics.
\newblock In S.~Emmerson, editor, {\em Music, Electronic Media and Culture},
  pages 194--216. Ashgate, Aldershot, 2000.

\bibitem{green2014nk}
O.~Green.
\newblock Nime, musicality and practice-led methods.
\newblock In {\em Proceedings of the 2014 conference on New Interfaces for
  Musical Expression (NIME)}, London, 2014.

\bibitem{Hayes:2011fk}
L.~Hayes.
\newblock Vibrotactile feedback-assisted performance.
\newblock In {\em Proceedings of the 2011 Conference on New Interfaces for
  Musical Expression}, Oslo, 2011. NIME.

\bibitem{Hayes2012kl}
L.~Hayes.
\newblock Performing articulation and expression through a haptic interface.
\newblock In ICMA, editor, {\em Proceedings of the 2012 International Computer
  Music Conference.}, Ljubljana, 2012.

\bibitem{Hayes:2013fk}
L.~Hayes.
\newblock Haptic augmentation of the hybrid piano.
\newblock In {\em Contemporary Music Review}, volume 32:5. Taylor and Francis,
  2013.

\bibitem{Merleau-Ponty:1962uq}
M.~Merleau-Ponty.
\newblock {\em Phenomenology of Perception (C. Smith, Trans.)}.
\newblock Routledge and Kegen Paul, 1962.

\bibitem{silephd}
S.~O'Modhrain.
\newblock {\em Playing by {F}eel: {I}ncorporating {H}aptic {F}eedback into
  {C}omputer-{B}ased {M}usical {I}nstruments}.
\newblock PhD thesis, Stanford University, CA, 2001.

\bibitem{Varela:1991fk}
F.~J. Varela, E.~Thompson, and E.~Rosch.
\newblock {\em The Embodied Mind}.
\newblock MIT, Cambridge, 1991.

\end{thebibliography}

\end{document}